\documentclass[twocolumn,showpacs,amsfonts,aps,prc,nofootinbib,floatfix,%
superscriptaddress]{revtex4}

\usepackage{amsmath}
\usepackage{bm}
\usepackage{graphicx}

\usepackage{epsfig}
\newcommand{\beq}{\begin{equation}}
\newcommand{\eeq}{\end{equation}}
\newcommand{\bea}{\vspace{0.25cm}\begin{eqnarray}}
\newcommand{\eea}{\end{eqnarray}}

\newcommand{\ro}{\mbox{{\boldmath
$\rho$}}}

\newcommand{\rb}{\mbox{{\bf
r}}}

\newcommand{\bb}{{{\bf b}}}

\newcommand{\E}{{{\bf E}}}
\newcommand{\J}{{{\bf J}}}
\newcommand{\Vb}{{{\bf V}}}

\newcommand{\B}{{{\bf B}}}

\def\lsim{\mathrel{\rlap{\lower4pt\hbox{\hskip1pt$\sim$}}
    \raise1pt\hbox{$<$}}}         
\def\gsim{\mathrel{\rlap{\lower4pt\hbox{\hskip1pt$\sim$}}
    \raise1pt\hbox{$>$}}}         

\newcommand{\landau}{L.D.~Landau Institute for Theoretical Physics,
        GSP-1, 117940, Kosygina Str. 2, 117334 Moscow, Russia}

\begin{document}


\title{
Electromagnetic response of 
quark-gluon plasma in heavy-ion collisions
}
\date{\today}

\author{B.G.~Zakharov}\affiliation{\landau}

\begin{abstract}
We study the electromagnetic response of the quark-gluon plasma 
in  $AA$-collisions at RHIC and LHC energies
for a realistic space-time evolution
of the plasma fireball.
We demonstrate that for a realistic 
electric conductivity the electromagnetic response 
of the plasma is in a quantum regime 
when the induced electric current 
does not generate a classical electromagnetic field,
and can only lead to a rare emission of single photons.
\end{abstract}
%

\maketitle


\section{Introduction}
Prediction of the chiral magnetic effect \cite{B1} in $AA$-collisions
stimulated studies of magnetic field generated in
heavy-ion collisions. 
In the noncentral $AA$-collisions 
the magnetic field perpendicular to the reaction plane
can reach the values 
$eB\sim 3m_{\pi}^{2}$ for RHIC ($\sqrt{s}=200$ GeV)
and a factor of 15 bigger  
for LHC ($\sqrt{s}=2.76$ TeV) conditions 
\cite{B1,Toneev_B1,Bzdak_B}.
In the initial stage 
the magnitude of the magnetic field falls rapidly with time 
($|B_{y}|\propto t^{-3}$, $y$-axis being perpendicular to the reaction plane). 
It was suggested \cite{Tuchin_review,Tuchin_B}  that the presence of the 
hot quark-gluon plasma (QGP)  may 
increase the lifetime of the strong magnetic field.
This may be  important for a variety of new
phenomena, such as the anomalous 
transport effects 
(for recent reviews, see \cite{Kharzeev_rev,Liao_rev}), 
the magnetohydrodynamics
effects \cite{Kharzeev-B,Hirano_ahydro}, the 
magnetic field induced photon production \cite{Kharzeev_phot,Tuchin_phot}.

The effect of the QGP on the evolution of the electromagnetic field
in $AA$-collisions has been estimated under the approximation
of a uniform static matter in
\cite{Tuchin_review,McL-B,Tuchin_B,Kharzeev-B}.
The difference between the calculations of
\cite{Tuchin_review,Tuchin_B,Kharzeev-B} and that of 
\cite{McL-B} is that 
in \cite{Tuchin_review,Tuchin_B,Kharzeev-B} the nuclei 
all the time move in the matter,
and in \cite{McL-B} it was assumed that the matter exists only
after the $AA$-collision at $t>0$.
In \cite{Tuchin_review,Tuchin_B,Kharzeev-B} 
a strong increase of the lifetime of the magnetic field in the presence
of the QGP was found. 
But one can expect that in the model of 
\cite{Tuchin_review,Tuchin_B,Kharzeev-B}
the matter effects should be overestimated, since 
there is an infinite time for the formation of the electromagnetic 
field around the colliding nuclei.  
In \cite{McL-B} 
it was obtained that for reasonable values of the conductivity
the matter does not increase the lifetime
of the strong ($eB/m_{\pi}^2\sim 1$) magnetic field, and
a significant effect was found only for the long-time
evolution where $eB/m_{\pi}^2\ll 1$.
The model of \cite{McL-B} seems to be more realistic, but
nevertheless it also may be too crude, since in reality
the matter does not occupy the whole space at $t>0$.
The plasma  fireball is formed
only in the region inside the light-cone $t>|z|$ between the flying apart 
remnants of the colliding nuclei, and in a restricted  transverse region 
of the overlap of the colliding nuclei.
Evidently, it is highly desirable to evaluate the electromagnetic 
response for a realistic space-time evolution of the matter. 
\begin{figure} [t]
\vspace{.7cm}
\begin{center}
\epsfig{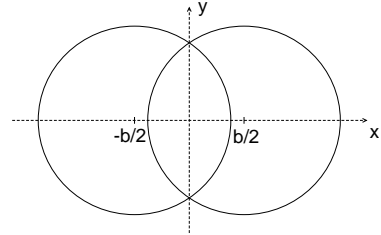}
\end{center}
\vspace{-0.5cm}
\caption[.]
{The transverse plane of a noncentral $AA$-collision with the impact
parameter $b$. 
}
\end{figure}

In this Letter we study the electromagnetic response 
of the QGP in the noncentral $AA$-collisions for a realistic
expanding plasma fireball which is created inside the light-cone $t>|z|$ 
in the almond-shaped transverse overlap of 
the colliding nuclei as shown in Fig.~1.
We demonstrate that the physical picture of the  
electromagnetic response is qualitatively different from the one
assumed in previous studies.
Our numerical results show 
that for a realistic electric 
conductivity the induced electromagnetic field generated in 
the fireball turns out to be too small for applicability of 
the classical treatment. We show
that for both RHIC and LHC energies the electromagnetic response
is essentially in the deep quantum regime
when one cannot talk about a classical electromagnetic field at all. 
In this regime the induced current in the QGP can just
produce single photons which freely leave the fireball 
without generation of an additional induced current in the QGP. 
The probability of the photon emission from this mechanism is very small,
and, due to a huge background from other mechanisms of the photon
production, an experimental observation of the photons from this
mechanism is practically impossible.

\section{Theoretical framework}
The electromagnetic field tensor satisfies the Maxwell equations 
\beq
\frac{\partial F_{\mu\nu}}{\partial x^{\lambda}}
+\frac{\partial F_{\nu\lambda}}{\partial x^{\mu}}
+\frac{\partial F_{\lambda\mu}}{\partial x^{\nu}}=0\,,
\label{eq:10}
\eeq
\beq
\frac{\partial F^{\mu\nu}}{\partial x^{\nu}}=-J^{\mu}\,.
\label{eq:20}
\eeq
For $AA$-collisions the current $J^{\mu}$ 
may be decomposed into two physically different pieces: 
\beq
J^{\mu}=J^{\mu}_{ext}+J^{\mu}_{in}\,.
\label{eq:30}
\eeq
Here  the term $J_{ext}^{\mu}$, which we call the external current,  is the
contribution of the fast right and left moving charged particles, 
which are mostly protons of the colliding nuclei. 
And the term $J^{\mu}_{in}$ is the induced current generated in the created 
hot QCD matter.
We decompose the field tensor 
also into the external and the induced pieces:
\beq
F^{\mu\nu}=F^{\mu\nu}_{ext}+F^{\mu\nu}_{in}\,.
\label{eq:40}
\eeq
Both $F^{\mu\nu}_{ext}$ and $F^{\mu\nu}_{in}$ 
separately satisfy the first Maxwell equation (\ref{eq:10}) and 
the following Maxwell equations with sources:
\beq
\frac{\partial F^{\mu\nu}_{ext}}{\partial x^{\nu}}=-J^{\mu}_{ext}\,,
\label{eq:50}
\eeq
\beq
\frac{\partial F^{\mu\nu}_{in}}{\partial x^{\nu}}=-J^{\mu}_{in}\,.
\label{eq:60}
\eeq
We assume that  Ohm's law is valid
in the fireball.
Then the induced current reads
\beq
J^{\mu}_{in}=\rho u^{\mu}+\sigma (F^{\mu\nu}_{ext}+F^{\mu\nu}_{in})u_{\nu}\,,
\label{eq:70}
\eeq
where $\sigma$ is the electric conductivity of the QCD matter, 
$\rho$ is its charge density, and 
$u^{\mu}$ is the four-velocity of the matter.
For the Bjorken 1+1D expansion \cite{Bjorken2} of the fireball
$u^{\mu}=(t/\tau,0,0,z/\tau)$, where $\tau=\sqrt{t^2-z^2}$ is the 
proper time.

The induced current (\ref{eq:70})
couples $F^{\mu\nu}_{in}$ 
to $F^{\mu\nu}_{ext}$.
And $F^{\mu\nu}_{ext}$ does not depend on the fireball 
evolution at all. 
We approximate $J^{\mu}_{ext}$ simply  
by the currents of the two colliding nuclei with the velocities
$\Vb_{R}=(0,0,V)$ and $\Vb_{L}=(0,0,-V)$ and with the impact parameters
$\bb_R=(0,-b/2)$ and $\bb_L=(0,b/2)$ as shown in Fig.~1.
We assume that in the center of mass frame of the $AA$-collision the 
trajectories of the centers of mass of the colliding nuclei 
in the longitudinal direction $z$ are $z_{R,L}=\pm Vt$.
The contribution of each nucleus to $F^{\mu\nu}_{ext}$ is given 
by the Lorentz transformation of its Coulomb field.
We write the electric and magnetic fields of a nucleus with the velocity 
$\Vb=(0,0,V)$ and the impact vector $\bb$ as 
\beq
\E_{T}(t,\ro,z)=\gamma \frac{E_{A}(r')(\ro-\bb)}{r'}\,,
\label{eq:80}
\eeq
\beq
E_{z}(t,\ro,z)=\frac{E_{A}(r')z'}{r'}\,,
\label{eq:90}
\eeq
\beq
\B(t,\ro,z)=[\Vb\times \E]\,.
\label{eq:100}
\eeq
Here $\gamma=1/\sqrt{1-V^{2}}$ is the Lorentz factor, 
$r'^{2}=(\ro-\bb)^{2}+z'^{2}$, $z'=\gamma(z-Vt)$,
and 
\beq
E_{A}(r)=\frac{1}{r^{2}}\int_{0}^{r}d\xi\xi^{2}\rho_{A}(\xi)
\label{eq:110}
\eeq
is the electric field of the nucleus in its rest frame, $\rho_{A}$ is
the nucleus charge density. In our calculations
we used for $\rho_{A}$ the Woods-Saxon parametrization.
From (\ref{eq:80})--(\ref{eq:110}) one can obtain that at 
$t^{2}\gsim (R_{A}^{2}-b^{2}/4)/\gamma^{2}$ (here $R_{A}$ is the 
nucleus radius, and $b$ is assumed to be $<2R_{A}$)
and $\rb=0$
the only nonzero $y$-component of the magnetic field 
for the two colliding nuclei is approximately
\beq
B_{y}(t,\rb=0)\approx \frac{\gamma Zeb}{4\pi(b^{2}/4+\gamma^{2}V^2t^{2})^{3/2}}\,.
\label{eq:120}
\eeq
From (\ref{eq:80})--(\ref{eq:110}) one can obtain that  
at $t\gg R_{A}/\gamma$    
$B_{y}(t,\ro,z=0)$ in the region $\rho\ll t\gamma$     
takes a simple $\rho$-independent form 
\beq
B_{y}(t,\ro,z=0)\approx
Zeb/4\pi\gamma^{2}t^{3}\,.
\label{eq:130}
\eeq
The quantity $R_{A}/\gamma$ is very small: $\sim 0.06$ 
for Au+Au collisions at RHIC energy $\sqrt{s}=200$ GeV,
and $\sim 0.004$ fm for Pb+Pb collisions at
LHC energy $\sqrt{s}=2.76$ TeV.
At $t^{2}\lsim (R_{A}^{2}-b^{2}/4)/\gamma^{2}$ the $t$-dependence of 
$B_{y}(\rb=0)$ flattens and at $t=0$ one can obtain
\beq 
B_{y}(t=0,\rb=0)\approx {\gamma Z eb}/{4\pi R_{A}^{3}}\,.
\label{eq:131}
\eeq
Here the right-hand side corresponds to the spherical nuclei.
For the realistic Woods-Saxon distribution of the protons
the result is just a bit ($\sim 5$\%) smaller.

\section{Model of the fireball}
The interaction of the Lorentz-contracted
nuclei lasts for a short time from $t\sim -R_{A}/\gamma$ to $R_{A}/\gamma$.
As the nuclei fly apart after the collision a hot 
fireball is created. 
It is widely accepted that the creation of the plasma fireball
goes through the thermalization of the glasma
longitudinal color fields created after multiple color exchanges
between the colliding nuclei.
We performed the calculations for the Bjorken longitudinal
expansion \cite{Bjorken2} of the fireball that gives the 
$\tau$-dependence of the entropy density $s\propto 1/\tau$.
We also performed the calculations accounting for the corrections
to the Bjorken picture from the transverse and the additional longitudinal
expansions of the fireball treating them perturbatively
as described in \cite{Ollitrault}. We assume that these corrections
come into play at $\tau_{0}=0.5$ fm. Roughly such $\tau_{0}$ is often used in 
the hydrodynamical simulations of $AA$-collisions
(for a recent review, see \cite{Heinz_flow}). 
But we observed that these corrections give a negligible effect.

For simplicity as in \cite{Ollitrault} we parametrize the initial entropy 
density profile at the proper time $\tau_{0}$ in a Gaussian form
\beq
s(x,y,\eta_{s})\propto \exp\left(-\frac{x^{2}}{2\sigma_{x}^{2}}
-\frac{y^{2}}{2\sigma_{y}^{2}}
-\frac{\eta_{s}^{2}}{2\sigma_{\eta}^{2}}\right)\,.
\label{eq:140}
\eeq
Here $\sigma_{x}$ and $\sigma_{y}$ are the root mean square widths of the
fireball in the transverse directions, and $\sigma_{\eta}$ is the root mean
square width in the space-time rapidity 
$\eta_{s}=\frac{1}{2}\ln\left(\frac{t+z}{t-z}\right)$.
We adjusted the parameters $\sigma_{x,y}(\tau_{0})$ using
the entropy distribution in the transverse coordinates at $\eta_{s}=0$
given by 
\beq
\frac{dS(\eta_s=0)}{d\eta_{s}d\ro}=
\frac{dS(\eta_s=0)}{d\eta_{s}}\cdot
\frac{\alpha \frac{dN_{part}}{d\ro}+(1-\alpha)\frac{dN_{coll}}{d\ro}}
{\alpha N_{part}+(1-\alpha)N_{coll}}\,,
\label{eq:150}
\eeq
where $dN_{part}/d\ro$ and $dN_{coll}/d\ro$  are the well known 
Glauber distributions of the participant nucleons and of the 
binary collisions (see, for instance, \cite{Glauber}).
We used in (\ref{eq:150}) $\alpha=0.95$. It  allows to reproduce well
the centrality dependence of the data on the pseudorapidity
density $dN_{ch}/d\eta$ from STAR \cite{N_STAR} for Au+Au collisions
at $\sqrt{s}=200$ GeV and from ALICE \cite{N_ALICE2} and CMS \cite{N_CMS}
for Pb+Pb collisions at $\sqrt{s}=2.76$ TeV. 
To fix the normalization of the entropy density 
we used the entropy/multiplicity ratio 
$dS/d\eta_{s}{\Big/}dN_{ch}/d\eta\approx 7.67$ obtained in \cite{BM-entropy}.
Making use of the data on $dN_{ch}/d\eta$ \cite{N_STAR,N_ALICE2,N_CMS} 
we obtained for the impact parameter $b=6$ fm
$\sigma_{x}(\tau_{0})\approx 2.3$ fm and $\sigma_{y}(\tau_{0})\approx 3.02$ fm
for RHIC and 
$\sigma_{x}(\tau_{0})\approx 2.42$ fm and $\sigma_{y}(\tau_{0})\approx 3.13$
fm for LHC.
We take
$\sigma_{\eta}(\tau_{0})\approx 2.63$ and $4.03$ for RHIC and 
LHC, respectively, that allow to reproduce qualitatively 
the experimental $\eta$-dependence of $dN_{ch}/d\eta$.
In evaluating the temperature through the entropy we used the ideal gas
formula for the number of flavors $N_f=2.5$.
It gives the temperature at the center of 
the fireball at $\tau=0.5$ fm: $T\approx 400$ MeV for RHIC and 
$T\approx 520$ MeV for LHC.

It seems likely that the model of the QGP as a conducting matter
makes sense at $\tau\gsim \tau_{0}$ when the hydrodynamics is
assumed to be applicable.
At present, the details of the thermal and chemical equilibration of the 
matter at early times $\tau\lsim \tau_{0}$ are unclear. 
Often it is assumed that the glasma  thermalization starts with the gluon
dominated stage and the production of quarks is somewhat delayed
(see, for instance, \cite{Leonidov_ph,BMueller_quarks,McLerran_quarks}).
But it is possible that immediately after the $AA$-collision 
the amount of quarks are close to that
for the chemically equilibrated QGP \cite{quark_CGC}. However,
even in this case it is hardly possible to describe the matter
in terms of the equilibrium conductivity because 
anyway the thermalization requires some time.
Nevertheless, since we would like to demonstrate that the 
electromagnetic response of the QGP is too small for
applicability of the classical treatment, 
we will consider a maximally optimistic scenario. We
assume that already at $\tau\gsim R_{A}/\gamma$ the conductivity makes sense
and equals to that for the equilibrium QGP with the entropy
density $\propto 1/\tau$, as in the Bjorken model \cite{Bjorken2}.
We solve the  Maxwell equations (\ref{eq:10}) (for $F^{\mu\nu}_{in}$) 
and (\ref{eq:60}) 
with the initial condition $F^{\mu\nu}_{in}=0$
at $\tau=R_{A}/\gamma$.

In our analysis we use the conductivity
obtained in the most recent lattice calculations for $N_f=3$ 
\cite{sigma_Amato} for $T\sim 140-350$ MeV. This analysis gives 
$\sigma/C_{em}T$ which rises smoothly from $\sim 0.07 $ at $T=150$ MeV
to $\sim 0.32$ at $T=350$ MeV. We parametrize the results of 
\cite{sigma_Amato} in the form
\beq
\sigma= C_{em}T f(T/1\,\text{GeV})\,,
\label{eq:160}
\eeq
\beq
 f(x) = 
\left\{ \begin{array}{ll}
    f_1  & \mbox{if $x \leq x_1$}\,,\\
 \frac{f_{1}(x_2-x)+f_{2}(x-x_1)}{x_{2}-x_{1}}
 & \mbox{if $x_1<x < x_2$}\,,\\
  \frac{f_2(x_3-x)+f_3(x-x_2)}{x_3-x_2}
 & \mbox{if $x \geq x_2 $}\,,
\end{array} \right.
\label{eq:170} 
\eeq
$f_{1-3}=0.0662$, $0.2153$, $0.3185$
and $x_{1-3}=0.1747$, $0.234$, $0.3516$.
The results of \cite{sigma_Amato} agree
qualitatively with that obtained within the Dyson-Schwinger equation approach
\cite{sigma_DSE}.

\section{Results and Discussion}
To numerically solve the Maxwell equations 
(\ref{eq:10}), (\ref{eq:60})
we rewrote them in the Milne coordinates $(\tau,x,y,\eta_{s})$  
which are convenient for imposing the initial condition at a given 
$\tau$ that we need. We used the Yee algorithm \cite{Yee}.
We found that it works well in the Milne coordinates as well.
We performed the calculations for the impact parameter $b=6$ fm
for Au+Au collisions at RHIC energy $\sqrt{s}=200$ GeV and for Pb+Pb
collisions at LHC energy $\sqrt{s}=2.76$ TeV.

\begin{figure} [t]
\vspace{.7cm}
\begin{center}
\epsfig{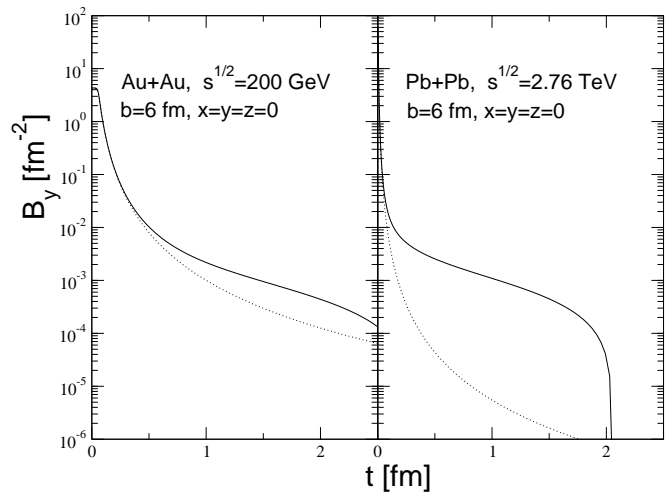}
\end{center}
\vspace{-0.5cm}
\caption[.]
{The time-dependence of magnetic field at $x=y=z=0$
for $AA$-collisions at $b=6$ fm at 
RHIC (left) and LHC (right) energies. Solid line: the total 
(external plus induced ) magnetic field; dotted line:
external magnetic field.
}
\end{figure}

In Fig.~2  we show the $t$-dependence of $B_{y}$ 
at $\rb=0$. And in Fig.~3  
we present the $x$-profile of $B_{y}$, $E_{x}$ and $E_{z}$ at $y=z=0$ for
$t=1$, $2$, and $4$ fm. We show separately the results for 
the total (external plus induced) and for the 
external fields. 
We present the curves for the Bjorken model. We observed 
that the corrections to the Bjorken model due to the transverse 
and the longitudinal expansion practically do not affect 
the electromagnetic response.
This is due to the fact that the conductivity is practically irrelevant
at $\tau\gsim 1$ fm where the corrections to the Bjorken model
can become noticeable.
We have checked this explicitly by performing the calculations for the 
conductivity switched off at $\tau>1$ fm. We
have found that this leads to a negligible change in 
the results.
To understand the effect of the self-interaction of the induced
component   $F_{in}^{\mu\nu}$ through its presence in the induced 
current on the right-hand side of (\ref{eq:70}) we also performed
the calculations neglecting $F_{in}^{\mu\nu}$ in
$J_{in}^{\mu}$. We found that this practically does not change the results.
It means that the induced field $F_{in}^{\mu\nu}$ is generated 
immediately after switching on of the conductivity
in the stage when the external field is still very large,
and its subsequent evolution goes practically as evolution of a free field 
in vacuum.
From Fig.~2 one can see that the induced magnetic field
becomes important at $t\gsim 0.5$ fm for RHIC and
at $t\gsim 0.1$ fm for LHC. At later times 
$B_{y}(\rb=0)$ becomes negative. As one can see
from Fig.~3 it is due to the development  of the typical spacial wave structure.
Fig.~3 shows that the induced magnetic and electric fields are
of the same order as it should be for a free electromagnetic field. 
\begin{figure} [t]
\vspace{.7cm}
\begin{center}
\epsfig{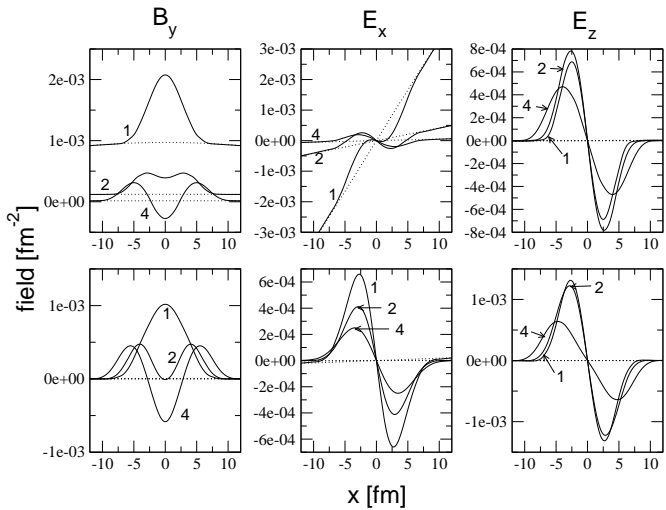}
\end{center}
\vspace{-0.5cm}
\caption[.]
{The $x$-dependence of 
$B_{y}$ (left), $E_{x}$ (middle) and $E_{z}$ (right) at $y=z=0$ for 
$t=1$, $2$, and $4$ fm (the curves marked by 1, 2 and 4)  
for Au+Au collisions at $\sqrt{s}=200$ GeV (upper) and 
Pb+Pb collisions at $\sqrt{s}=2.76$ TeV (lower) 
for the impact parameter $b=6$ fm. Solid line: the total 
(external plus induced ) fields; dotted line:
external fields.
}
\end{figure}

Our results for the magnetic field shown in Fig.~2 are considerably 
smaller than that obtained  in the model of a uniform matter 
existing all the time in \cite{Tuchin_B} 
(for Au+Au collisions at $\sqrt{s}=200$ and $b=7$ fm) and in \cite{Kharzeev-B}
(for Pb+Pb collisions at $\sqrt{s}=2.76$ TeV $b=7$ fm).
At $t=1$ fm our predictions are smaller than that of \cite{Tuchin_B}
by a factor of $\sim 50$, and for \cite{Kharzeev-B} by a factor
of $\sim 90$ (in obtaining these numbers we have taken into account
that the magnetic field is approximately proportional to $b$,
and rescaled the results of \cite{Tuchin_B,Kharzeev-B} by the factor $6/7$).
Unfortunately, there is some problem with comparing of our
results with that 
of \cite{McL-B} (for Au+Au collisions at $\sqrt{s}=200$ and 
$b=6$ fm). By examining the external magnetic field 
shown in Fig.~1 of \cite{McL-B} 
we have found that prediction for the external field given there
is clearly wrong. Indeed, from (\ref{eq:131}) one obtains 
 $eB_{y}(t=0,\rb=0)/m_{\pi}^{2}\approx 2.73$, and accurate calculations
with the Woods-Saxon density give for this quantity a bit smaller
value $2.6$, while
Fig.~1 of \cite{McL-B}  gives $eB_{y}(t=0,\rb=0)/m_{\pi}^{2}\approx 7$.
Thus at $t=0$ \cite{McL-B} overestimates the field by a factor of $\sim 2.7$.  
The $t$-dependence of the external field in \cite{McL-B}
is also wrong, say Fig.~1 gives $eB_{y}(t=R_A,\rb=0)/m_{\pi}^{2}
\approx1.8\cdot 10^{-4}$, while formula (\ref{eq:130})
gives for this quantity a value smaller by a factor of $\sim 80$. 
It is possible that the above strange behavior of the external
field in Fig.~1 of \cite{McL-B} is just a consequence of some errors
in axis variables. 
We will assume that it is really the case. Then using the correct value 
$eB_{y}(t=0,\rb=0)/m_{\pi}^{2}\approx 2.6$ for normalization
of the total (external plus induced) field in the region 
where in Fig.~1 of \cite{McL-B} it flattens, we obtain
$eB_{y}(t\sim 1 \mbox{fm},\rb=0)/m_{\pi}^{2}\sim 0.017$. 
It is about a factor of $3.7$ smaller than prediction 
of \cite{Tuchin_B} (rescaled by a factor $6/7$ accounting
for difference in $b$), and by 
a factor of $13$ bigger than our prediction 
$eB_{y}(t\sim 1 \mbox{fm},\rb=0)/m_{\pi}^{2}\approx 0.0013$.

The magnitude of the difference in predictions of \cite{Tuchin_B} and 
\cite{McL-B} seems to be quite reasonable since in \cite{Tuchin_B}
there is a contribution from the unrestricted region 
at $t<0$. But the difference between
\cite{McL-B} and our results seems to be too big to be explained by the
difference in the space-time distribution of the conductivity.
One can expect that the latter could only give a factor of $\sim 2-4$.
So there must be a mechanism which enhances the medium effect
in the model of \cite{McL-B} as compared to our one.
It seems likely that it is the difference in the induced current 
for the static matter and the matter with the
Bjorken longitudinal expansion. It can be seen by comparing 
the form of the induced current in these two models.
Our calculations show that the effect of the induced electromagnetic
field in the induced current is practically negligible. So 
we can consider in the induced current (\ref{eq:70})
only the term with $F_{ext}^{\mu\nu}$. 
For the static matter the dominating transverse component
of the current  for each of the colliding 
nuclei reads $\J_{T}=\sigma \E_{T}$. So each nucleus produces a running
pancake-like distribution of the induced current. It acts as
an antenna radiating the induced electromagnetic field.
One can easily show that for the
Bjorken expansion of the matter the $\E_{T}$ in the current is replaced
by the transverse electric field in the comoving frame. The latter is
suppressed by a factor $\exp{(-|\eta_{s}|)}$ as compared to $\E_{T}$ in the
center mass frame. In the vicinity of the nuclei this factor may be
$\sim 1/\gamma$. Of course, the induced field acquires the contributions
radiated from the points with different $\eta_{s}$, and the resulting 
suppression factor should be bigger than $1/\gamma$. Nevertheless,
it is clear that the suppression effect may be quite strong.
The finite size of the fireball 
in the rapidity also can give an additional suppression of the
induced field. Unfortunately, the induced field
for the conditions of Ref. \cite{McL-B} cannot be computed directly 
with our code because it is written in the Milne coordinates and works only
inside the region $t>|z|$ (for finite values of the proper time 
$\tau=\sqrt{t^2-z^2}$). However, we tested that in our formulation
for the finite fireball with zeroth longitudinal velocity and a flat
distribution of the conductivity in the rapidity the induced field
is really enhanced by a factor of $\sim 8$ and the results
turns out to be qualitatively similar to that of \cite{McL-B}.
In the formulation of \cite{Tuchin_B,Kharzeev-B}, where an unrestricted 
region of time and transverse space is involved into the formation
of the induced field at a given space-time point, the enhancement 
may be considerably bigger. Therefore the observed disagreement of
our results with that of \cite{Tuchin_B,Kharzeev-B} do not seem
to be unrealistic.

From the results shown in Figs.~2,~3 one can show
that the electromagnetic response of the plasma fireball
is in reality in a quantum regime.
Indeed, it is known \cite{LL4} that 
the classical treatment of an electromagnetic field 
is valid when
\beq
|\E|,\,|\B|\gg 1/\Delta t^{2}\,,
\label{eq:180}
\eeq
where $\Delta t$ is the typical time of the observation. For
$\Delta t$ one can simply take the typical time of the variation
of the fields \cite{LL4}. The inequality (\ref{eq:180}) follows
from the condition that for the classical description 
the occupation numbers should be large.
It is especially transparent for a free field occupying (at a given instant)
a restricted region of space. If the size of the region is $L$, then 
$\Delta t\sim L$, and the dominating Fourier component should have 
a frequency $\omega\gsim 1/\Delta t$. Then by requiring that 
the energy of the field, which is $\sim L^{3}(\E^{2}+\B^{2})/2$, is much
bigger than the typical one photon energy $\omega$ one obtains (\ref{eq:180}).
In our case one can take $\Delta t\sim t$. 
From the curves shown in Figs.~2,~3 one can easily see 
that the induced fields are much smaller than 
$1/t^{2}$. It means that the typical photon occupations numbers are much
smaller than unity. In this situation one cannot talk about classical
fields, and the electromagnetic response of the QGP should be described
as radiation of single photons. This fact is quite evident from
the $x$-profile of the magnetic and electric fields shown in Fig.~3. 
On the one hand,
one can see that the typical wave vector is of the order of $1$ fm$^{-1}$. 
On the other hand, if we estimate the total energy of the field 
$U$, say,  at $t\sim 4$ fm taking (with a large excess) 
for the volume $V\sim 1000$ fm$^{3}$ 
we obtain in fm units $U\lsim 0.01$ fm$^{-1}$. It is much 
smaller than the expected typical photon energy $\sim 1$ fm$^{-1}$.
This means that we have a situation of the deep quantum regime when
the electromagnetic response of the QGP is a very rare emission of 
the single photons which freely leave the fireball. The fact that
the photons are not absorbed in the fireball is evident from
calculation of the photon attenuation length $l_{a}\approx 2/\sigma$, which,
in our case, turns out to be very large $l_{a}\sim 100$ fm for $T\sim 250$ MeV.
This is why we have found the negligible effect of switching off 
the conductivity at $\tau>1$ fm.
An experimental observation of the photons radiated by the induced current
is practically impossible due to a huge background from other
mechanisms of the photon production.
Note that since the electromagnetic response of the QGP
consists of emission of the single photons, it
cannot contribute to the magnetohydrodynamics
effects \cite{Kharzeev-B,Hirano_ahydro} and to the 
magnetic field induced photon production \cite{Kharzeev_phot,Tuchin_phot}.

One remark is appropriate at this point. 
From Figs.~2,~3 (or from Eq. (\ref{eq:130})) one can see that the external 
fields also do not satisfy the criterion (\ref{eq:180}).
It seems to be in contradiction with the estimate of the 
photon occupation numbers for the electromagnetic field of a fast
nucleus. Indeed, one can easily show that for $\gamma\gg 1$
the typical occupation numbers are 
\beq
N\sim Z^{2}\alpha_{em}/4\pi\,.
\label{eq:181}
\eeq
From (\ref{eq:181}) one sees that 
for $Z\sim 100$ the classical approximation should work well, at least 
except for the tail regions (in the longitudinal direction) 
where the field becomes very small and the situation may be a quantum one.
This puzzling situation with the contradiction of the criterion
(\ref{eq:180}) and the estimate (\ref{eq:181}) is related to the fact that 
in \cite{LL4}
in deriving (\ref{eq:180}) it was implicitly 
assumed that the distribution of the photon modes is more or less
isotropic. This is clearly not true for the field of a fast nuclei,
when the field has a pancake-like form and the modes 
are strongly collimated in the direction of the nucleus velocity.
One can easily show that in this situation the criterion (\ref{eq:180})
should be replaced by 
\beq
|\E|,\,|\B|\gg 1/\Delta t \Delta \rho\,,
\label{eq:182}
\eeq
where $\Delta \rho$ is the typical scale of variation of the fields
in the transverse directions. One can see that, taking 
$\Delta \rho\sim \rho$, where $\rho$ is the transverse distance from the 
nucleus, and  $\Delta t\sim \rho/\gamma$, the criterion (\ref{eq:182}) 
and the estimate (\ref{eq:181}) of the occupation numbers give the same
condition for the validity of the classical description.
For the induced field which occupies the whole fireball region
and is not collimated in one direction the criterion (\ref{eq:180})
should work. Our estimate of the occupation numbers based on the 
comparison of the energy with the typical wave vector confirms this. 
We would like to emphasize that, in any case, our conclusion
about the quantum character of the electromagnetic response is 
completely independent of the situation (classical or quantum) 
with the external field at later times where we apply our energy
argument because in this region the external field becomes very small
as compared to the induced field
and can simply be ignored.

Note that our main result,  that the electromagnetic response of the
fireball is in a quantum regime, persists for a wide range of the
electric conductivity. 
We checked that for $\sigma=7C_{em}T$ from the early analysis 
\cite{sigma_Gupta}, which is by a factor of $\sim 20$ (at $T\sim 300$ MeV)
larger than that of \cite{sigma_Amato}, 
the induced magnetic and electric fields 
also violate the inequality (\ref{eq:180}).
It is hardly possible that $\sigma$ can be bigger than that
of \cite{sigma_Gupta}, since even it seems to be
too large. 
Indeed, using the Drude formula one can show that,
in terms of the quark collisional time $\tau_{c}$,
$\sigma$ from \cite{sigma_Gupta} corresponds
to $\tau_{c}/\tau\sim 10-20$ for $\tau\sim 0.5-1$ fm
and $\tau_{c}/\tau\sim 4-8$ for $\tau\sim 2-4$ fm.
Such large ratios say that the quarks are in a ballistic regime.
This cannot be reconciled with the successful
hydrodynamical description of the flow effects in $AA$-collisions 
\cite{Heinz_flow}.
The results of \cite{sigma_Amato} seem to be more realistic.
In this case
$\tau_{c}/\tau\sim 0.2-0.3$ for $\tau\sim 0.5-1$ fm
and $\tau_{c}/\tau\sim 0.05-0.1$ for $\tau\sim 2-4$ fm
which look quite reasonable (from the viewpoint of the 
applicability of the hydrodynamics and of the 
model of a conducting matter).

In summary, we have studied the electromagnetic response of the 
QGP in the noncentral $AA$-collisions at RHIC and LHC energies
by solving the Maxwell equations for a realistic 
space-time evolution of the plasma fireball. 
We demonstrate that the resulting induced electromagnetic field 
turns out to be too small for applicability of the classical treatment,
and in reality the electromagnetic response is in a quantum regime,
when the induced electric current in the plasma fireball 
cannot generate a classical electromagnetic field at all.
In this regime the electromagnetic response consists only of a rare emission of 
the single photons. Thus, the emerging physical picture of the electromagnetic
response of the QGP differs qualitatively from that assumed
previously.

\begin{acknowledgments}
I thank the anonymous referees for their stimulating comments
that helped to improve the paper.
This work is supported 
in part by the 
grant RFBR
12-02-00063-a
and the program SS-3139.2014.2.
\end{acknowledgments}

\end{document}